\documentclass{CVIS}
\usepackage{amsmath,bm}
\usepackage{graphicx}
\usepackage{dcolumn}
\usepackage{subcaption,soul}
\usepackage{caption}
\usepackage[numbers,sort&compress,sectionbib]{natbib}
\usepackage[pagebackref=flase,breaklinks=true,letterpaper=true,colorlinks=true,linkcolor = blue, urlcolor = black,citecolor=black,bookmarks=true]{hyperref}
\usepackage{url}
\usepackage{mleftright}
\usepackage{multirow}

\usepackage{titlesec}
\usepackage[redeflists]{IEEEtrantools}

\usepackage{tikz}
\usetikzlibrary{positioning,shapes,arrows}
\usepackage{booktabs}

\title{A Trustworthy Framework for Medical Image Analysis with Deep Learning}

\begin{document}
\bstctlcite{IEEEexample:BSTcontrol}
\sloppy
\author{
\begin{tabularx}{\textwidth}{X X}
Kai Ma & University of Waterloo \\
Siyuan He  & University of Waterloo \\
Pengcheng Xi & National Research Council Canada \\
Ashkan Ebadi & National Research Council Canada\\
Stephane Tremblay & National Research Council Canada \\
Alexander Wong & University of Waterloo \\
\multicolumn{2}{l}{Email: \{k78ma, sy4he, a28wong\}@uwaterloo.ca, \{pengcheng.xi, ashkan.ebadi, stephane.tremblay\}@nrc-cnrc.gc.ca}
\end{tabularx}
}

\maketitle
\begin{abstract}
Computer vision and machine learning are playing an increasingly important role in computer-assisted diagnosis; however, the application of deep learning to medical imaging has challenges in data availability and data imbalance, and it is especially important that models for medical imaging are built to be trustworthy. Therefore, we propose TRUDLMIA, a trustworthy deep learning framework for medical image analysis, which adopts a modular design, leverages self-supervised pre-training, and utilizes a novel surrogate loss function. Experimental evaluations indicate that models generated from the framework are both trustworthy and high-performing. It is anticipated that the framework will support researchers and clinicians in advancing the use of deep learning for dealing with public health crises including COVID-19.
\end{abstract}

\section{Introduction}

The COVID-19 pandemic continues to affect lives around the world. Medical imaging, including chest X-rays, plays a key role in diagnosis. Using computer vision and deep learning, computer-assisted diagnosis has shown great potential in this field \cite{DESAI2020100013, Rehouma2021}. Moreover, improving on the task of COVID-19 chest X-ray classification may lead to advancements for similar tasks. However, three main issues have been identified in this field. \\

\textbf{Limited data.} Datasets for medical imaging, especially for novel diseases such as COVID-19, are typically small compared to natural image datasets, making model training more difficult. As such, medical feature learning is commonly conducted through transfer learning \cite{bigTransfer}, with large-scale supervised learning followed by down-stream fine-tuning. This approach, however, is limited by the relevance of the large-scale data to the downstream task, as well as label quality. Consequently, self-supervised learning (SSL) has been proposed and has shown comparable performance to state-of-the-art supervised models \cite{simclr, moco}. The methodology behind \textit{contrastive} self-supervised learning, which aims to learn feature representations by comparing closely related data samples to each other, is especially relevant, as medical images are extremely similar and can appear identical to untrained eyes. \\

\textbf{Class imbalance.} Aside from small datasets, another important issue for medical imaging is class imbalance, where there are significantly more negative (benign) data samples than positive (malignant) ones. Thus, models are heavily biased toward the majority negative class and exhibit poor predictive performance for the minority positive class, which is often more important for medical diagnosis. A way to combat this during the training process is to maximize the AUC (area under the ROC curve) instead of minimizing cross-entropy (CE) loss \cite{aucrobust}. This is suitable for imbalanced data, as maximizing AUC aims to rank the prediction score of positive samples higher than negative ones. However, AUC maximization is more sensitive to model changes, making it less practical than minimizing CE loss \cite{aucrobust}. \\



\textbf{Low trustworthiness.} In the context of medical AI, trustworthiness of predictions is important to both patients and clinicians. An existing problem is that deep neural networks optimized with the standard CE loss function tend to be overly cautious for the minority class, while being overconfident for the majority class \cite{ghost}. This problem is especially hard to deal with, as model trust quantification is a relatively new and undeveloped area. Existing literature typically focuses on evaluating the trust for a prediction from a single data sample \cite{uncertestimation, uncertprop}; therefore, these approaches often suffer from various weaknesses, including low interpretability and being limited to Bayesian networks \cite{uncertbayesian}. To deal with these limitations, a concept of ``question-answer" trust has been introduced \cite{trustwong}, where the trustworthiness of a model is determined by its behaviour when answering questions, such that undeserved confidence is penalized while well-placed confidence is rewarded. Through this method, a simple scalar ``trust score" is introduced, such that a higher trust score indicates a more trustworthy model. To remedy the problem of low model trust, we use Deep AUC Maximization to replace traditional CE loss with the robust AUC min-max margin loss \cite{aucrobust}. 
\\



To address the aforementioned issues, we propose \textbf{TRUDLMIA}, a \textbf{simple and trustworthy deep learning framework} for medical image analysis. Both supervised and self-supervised learning are combined for effective medical image feature learning and a novel surrogate loss function is adopted to build high-performing, high-trust models. The framework adopts a model-agnostic modular design for generalization capabilities. \\



In summary, our contributions and findings are three-fold: 
\begin{itemize}
    \item We propose a general deep learning framework for medical image analysis which can be used to build high-performing, high-trust models;
    \item We show that fine-tuned models with self-supervised pre-training surpass supervised ones for COVID-19 classification, including state-of-the-art deep learning models designed specifically for the task;
    \item AUC maximization with margin loss leads to more effective feature learning and higher trustworthiness, effectively dealing with the problems of class imbalance and prediction under/over-confidence.
\end{itemize}

\section{Literature Review}
\label{sec:litreview}
In computer vision, SSL has gained popularity for learning representations, requiring no labels unlike supervised training. The SSL approaches can be categorized into generative or contrastive/discriminative. Two mainstream contrastive approaches, \textit{momentum contrast} (MoCo) \cite{moco} and \textit{SimCLR} \cite{simclr}, learn features by comparing data samples to each other. MoCo approaches this task through a mechanism analogous to dictionary look-up. Through a contrastive loss function, a visual representation encoder is trained by matching encoded queries to a dictionary of encoded keys. SimCLR, however, focuses on the use of data augmentations on the same image. The SimCLR aims to minimize the distance between data augmentations of the same image, while maximizing the distance between different images. 
\\

Contrastive learning methods have been shown to be effective in medical contexts. The MoCo model has been trained on a chest X-ray data-set to produce the MoCo-CXR model \cite{moco-cxr}. Subsequent fine-tuning experiments show that models initialized with MoCo-CXR outperformed non-MoCo-CXR counterparts, especially on limited training data. Experiment on a dataset that is unseen during pre-training also shows that MoCo-CXR pre-training has good transferability across chest X-ray datasets and tasks.  \\

In dealing with the COVID-19, the SSL methods exhibit advantages over supervised counterparts because of limited data with labels. In \cite{tompaper}, the authors showed results that self-supervised pre-training using MoCo led to better results than supervised pre-training for screening COVID-19 patients. In \cite{https://doi.org/10.48550/arxiv.2101.04909}, the authors pre-train the MoCo model on chest X-rays to learn more general image representations to use for prognosis tasks, differing from previous work in that existing solutions leverage supervised pre-training on non-COVID images, an approach limited by the difference between the pre-training data and the target COVID-19 patient data. It thus achieves comparable prediction accuracy to that of experienced radiologists analysing the same information. SimCLR has also been applied for medical AI. In \cite{https://doi.org/10.48550/arxiv.2101.05224}, the authors propose multi-instance contrastive learning, a novel approach that generalizes contrastive learning to leverage special characteristics of medical image analysis. They observe that SSL pre-training on ImageNet, followed by additional pre-training on unlabeled domain-specific medical images, improves classifier accuracy \cite{https://doi.org/10.48550/arxiv.2101.05224}. \\

The quantification of model trustworthiness is a new and undeveloped area, compared to other deep learning performance metrics. Existing work typically focuses on evaluating trustworthiness for a prediction made on a single data sample, either through measuring agreement with a nearest-neighbour classifier \cite{nearestneighour, xiong2022birds} or estimations for model uncertainty \cite{uncertprop, uncertbayesian}. Other newer approaches employ complex frameworks such as probabilistic descriptions based on network topologies \cite{deeptrust}, and even cloud-based heuristics that rely on a large number of models \cite{cloudtrust}. However, these approaches often suffer from severe weaknesses --- their trust quantification is often highly complex, hard to interpret, or limited to certain neural networks like Bayesian ones \cite{uncertbayesian}. To deal with these limitations, a concept of ``question-answer" trust has been introduced in \cite{trustwong}, where the trustworthiness of a model is determined by its behaviour when answering questions correctly or incorrectly. Consequently, undeserved confidence is appropriately penalized while well-placed confidence is rewarded. Through this method, a simple scalar ``trust score" is introduced to express question-answer to practice. 
\\


In medical imaging, AUC score has become a common metric to compare deep learning methods, and directly maximizing AUC score is a proven method of improving model performance. Furthermore, proponents of AUC maximization claim that it is optimal for handling imbalanced data, especially for tasks where the positive class is important, since maximizing AUC aims to rank the prediction score of any positive data higher than any negative data \cite{aucrobust}. An ongoing area of study is the design of surrogate loss functions for AUC maximization, with the standard naive approach being a simple pairwise surrogate loss based on the definition of the AUC score \cite{pairwise}. However, this suffers from severe scalability issues and sensitivity to noise. To alleviate this, the authors of \cite{aucrobust} propose AUC min-max margin loss, a novel surrogate loss function for maximizing AUC score, which uses a squared hinge function (common in margin-based SVM classifiers). Deep AUC Maximization (DAM) with AUC min-max margin loss has shown state-of-the-art results on various difficult tasks with unbalanced data, including medical imaging tasks such as CheXpert \cite{chexpert} and melanoma detection. \\


\section{Methodology}

\subsection{Framework Design}
\begin{figure}[ht]
	\begin{center}
		\includegraphics[scale = 0.38]{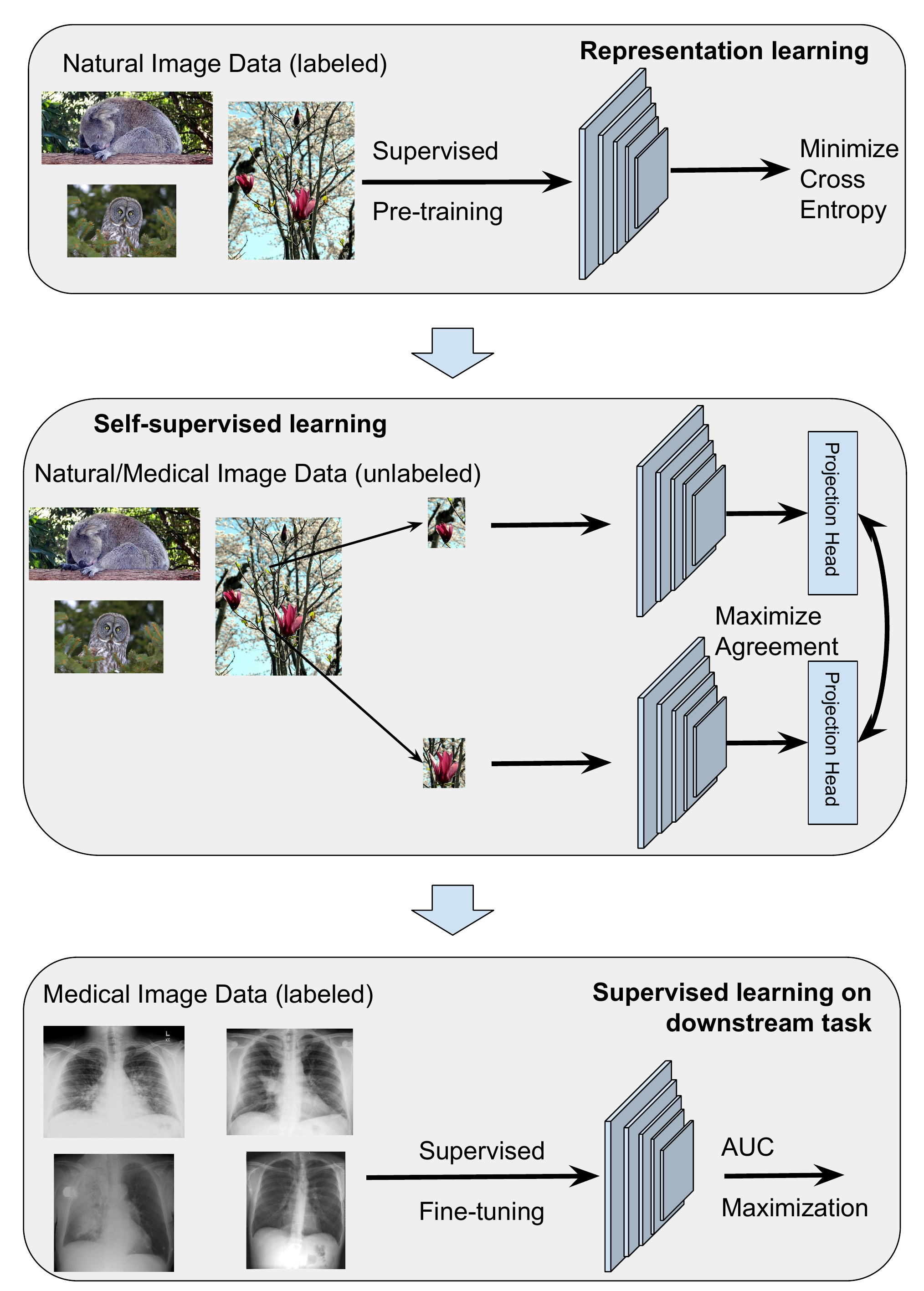}
	\end{center}
	\caption{Overview of the proposed deep learning framework.}
	\label{fig:framework}
	\label{fig1}
\end{figure}

The proposed TRUDLMIA framework comprises three main modules: i) generic learning through large-scale supervised pre-training using natural images (ImageNet \cite{imagenet}), ii) adapted learning through large-scale self-supervised learning using natural images or domain images without labels, and iii) targeted learning through supervised fine-tuning on downstream tasks using a labeled dataset (see Fig. \ref{fig:framework}). Conducting supervised pre-training (module i) before self-supervised pre-training (module ii) means less epochs of the latter are required, making the framework much more computationally efficient as SSL can be very costly. The three modules are built upon each other in order to build trustworthy models with optimal performance, despite the fact that training data on target tasks are limited in quantity and imbalanced. For validation, the model achieving the best validation accuracy is saved and evaluated on the unseen test split.\\


We study two main-stream SSL approaches, namely SimCLR and MoCo, on their performance and trustworthiness as pre-training for our framework. In the comparison below and ablation studies, we mainly use SimCLR, as we found that it performed better than MoCo for the given task (results shown in Section \ref{plugin}). We use AUC maximization with AUC min-max margin loss in module iii) of the TRUDLMIA framework, which has several benefits over traditional cross-entropy (CE) loss. Most importantly, AUC maximization is better at handling imbalanced data \cite{aucrobust}, being more resistant to trust issues caused by CE loss (over-confidence in the majority class and over-cautiousness in the minority class \cite{ghost}). \\

The TRUDLMIA framework adopts a plug-in architecture in the computation of image features. In our study, we adopt main-stream deep convolutional neural network (CNN) models, i.e., ResNet \cite{https://doi.org/10.48550/arxiv.1608.06993} and DenseNet \cite{https://doi.org/10.48550/arxiv.1512.03385}, which can be replaced with other network architectures. The two self-supervised learning (SSL) approaches compared in module ii) are also replaceable. Likewise, the AUC maximization used in the module iii) can be replaced with alternative loss functions.

\subsection{Trust Score Computation} \label{subsec:trustcomp}
We compute trust scores for models based on a method introduced in \cite{trustwong}. Given a question $x$, an answer $y$ with respect to a model $M$, such that $y = M(x)$, and $z$ representing the correct answer to $x$, we then use $ R_{y=z \mid  M}$ to denote the space of all questions where the answer $y$ given by model $M$ matches the correct answer $z$. Likewise, we use $R_{y \neq z \mid  M}$ to denote the space of all questions where the answer $y$ given by model does not match the correct answer. We also define the confidence of $M$ in an answer $y$ to question $x$ as $C(y\mid x)$. Thus, \textit{question-answer trust} of an answer $y$ given by model $M$ of a question $x$, with knowledge of the correct answer $z$, is defined as
\[  Q_z(x,y) =  \left\{
\begin{array}{ll}
      C(y\mid x)^\alpha, & \text{if } x \in R_{y=z \mid M} \\
      (1 - C(y\mid x))^\beta, & \text{if } x \in R_{y\neq z \mid M}, \\
\end{array}
\right. \]
with $\alpha$ and $\beta$ denoting reward and penalty relaxation coefficients. \\

To integrate the trust score computation into the models trained through our framework, we first calculate an optimal threshold value by maximizing F1-score on a validation split. Data samples are classified using the threshold, and outputs are then normalized such that negative predictions are scaled between 0 and 0.5 and positive predictions are scaled between 0.5 and 1. This allows us to express model confidence, which is then used to compute the \textit{trust score} according to the question-answer method introduced above. We use $\alpha = 1$ and $\beta = 1$, equally rewarding well-placed confidence and undeserved overconfidence. This is done for all of the positive samples in the unseen test split. Finally, an overall positive class trust score for the model is determined by calculating the mean of the computed individual scores.

\section{Experimental Results}

\subsection{Dataset}
\begin{table}[ht]
\centering
\caption{Data split for COVIDx 8B}\label{tab:data}
\begin{tabular}{c  c  c  c}
\toprule
\textbf{Split} & \textbf{Negative}  & \textbf{Positive} & \textbf{Total}\\
\midrule
Train    & 13,793   & 2,158  & 15,951  \\
Test    & 200   & 200  & 400  \\
\bottomrule
\end{tabular}
\end{table} 


Various datasets are used in TRUDLMIA modules. In supervised pre-training (module i), the deep CNN models are pre-trained on the ImageNet \cite{imagenet} dataset. In self-supervised learning (module ii), the MoCo model is pre-trained on both ImageNet and MIMIC-CXR dataset \cite{bib2}. The SimCLR model is pre-trained on the ImageNet dataset. For the downstream task (module iii), the COVIDx dataset (Version 8B) \cite{covidnetlinda}, a small dataset with a high class imbalance, is used for fine tuning the models end-to-end. The dataset training/testing split is shown in Table \ref{tab:data}). All models built in our experiments are evaluated on the same test subset. 


\subsection{COVID-CXR-SSL}
Our top-performing model, dubbed COVID-CXR-SSL, demonstrates both high performance and trust score (see Table \ref{optimalModels}). The model uses ResNet in module i) and SimCLR in module ii), followed by fine tuning on the COVIDx dataset in module iii). \\




We compare the COVID-CXR-SSL with COVID-Net CXR-2 \cite{cxr-2} and COVID-Net CXR-3 \cite{medusa}, high-performing models that have been designed for the same dataset. COVID-Net CXR-2 uses machine driven design to automatically discover highly customized macro/micro-architecture designs. COVID-Net CXR-3 employs a self-attention mechanism (MEDUSA). Both models are state-of-the-art, consistently surpassing high-performing models on various medical imaging tasks \cite{medusa}. 
Our COVID-CXR-SSL model outperforms both COVID-Net CXR-2 and COVID-Net CXR-3 across all metrics on COVIDx V8B. It is noted that the COVID-Net CXR models use input images at a resolution of $480\times480$, while our model uses a resolution of only $224\times224$. \\

\begin{table}[ht]
\caption{Model performance and trust scores for COVID-Net CXR-2, COVID-Net CXR-3 and COVID-CXR-SSL}\label{optimalModels}
\begin{center}
\begin{minipage}{\columnwidth}
\begin{tabular*}{\textwidth}{@{\extracolsep{\fill}}lccccc@{\extracolsep{\fill}}}
\toprule%
{Model} & \multicolumn{2}{c}{Precision} & \multicolumn{2}{c}{Sensitivity} & \multicolumn{1}{c}{Trust} \\\cmidrule{2-3}\cmidrule{4-5}%
 & Pos. & Neg. & Pos. & Neg.  & \\
\midrule
COVID-Net CXR-2 & 0.970 & 0.955 & 0.956 & 0.970 & - \\
COVID-Net CXR-3 & 0.990   & 0.975  & 0.975   & 0.990 & -   \\
COVID-CXR-SSL  & \textbf{1.000}   &  \textbf{0.980} & \textbf{0.980}   & \textbf{1.000} & \textbf{0.964} \\
\bottomrule
\end{tabular*}
\end{minipage}
\end{center}
\end{table}

\subsection{Ablation Study}

We conduct an ablation study to investigate the contribution of different modules in the TRUDLMIA framework. We start with fine-tuning a pre-trained ResNet on the COVIDx dataset as a baseline (model ``SL"). The pre-trained ResNet model is also used as a backbone architecture for training with the SimCLR architecture followed by fine-tuning it using the CE loss function (model ``SL+SSL"). Furthermore, the SimCLR model is also fine-tuned using the AUC maximization loss function (model ``SL+SSL+AUC"). Both the ``SL+SSL" and ``SL+SSL+AUC" models are fine tuned for 200 epochs. Table \ref{ablationModels} lists the performance metrics and trust scores computed on the models. We obtain an increase of about $6\%$ on precision and sensitivity metrics and $4\%$ on trust score from the adoption of SSL. AUC maximization further improves the performance while improving trust score slightly. \\

\begin{table}[ht]
\caption{Ablation study on model performance and trust scores for different model architectures}\label{ablationModels}
\begin{center}
\begin{minipage}{\columnwidth}
\begin{tabular*}{\textwidth}{@{\extracolsep{\fill}}lccccc@{\extracolsep{\fill}}}
\toprule%
{Architecture} & \multicolumn{2}{c}{Precision} & \multicolumn{2}{c}{Sensitivity} & \multicolumn{1}{c}{Trust} \\\cmidrule{2-3}\cmidrule{4-5}%
 & Pos. & Neg. & Pos. & Neg.  & \\
\midrule
SL  & 1.000   & 0.885  & 0.870   & 1.000 & 0.918   \\
SL+SSL  & 1.000  &  0.939 & 0.935   & 1.000 & 0.952 \\
SL+SSL+AUC  & \textbf{1.000}   &  \textbf{0.952} & \textbf{0.950}   & \textbf{1.000} & \textbf{0.954} \\
\bottomrule
\end{tabular*}
\end{minipage}
\end{center}
\end{table}

\subsection{Selection of SSL Plug-in}
\label{plugin}
Given the choice of different SSL approaches, we conduct a comparison with MoCo architecture in the TRUDLMIA framework. After fine-tuning for 100 epochs, we select the best models for evaluation and provide their performance metrics in Table \ref{sslModels}. The SimCLR based model outperforms the MoCo-based one across all metrics. Our results also indicate that pre-training with natural images on ImageNet is more effective than pre-training on large-scale medical image dataset, MIMIC-CXR, despite the latter being more relevant to the downstream task. \cite{bib2}.

\begin{table}[ht]
\caption{Model performance and trust scores for different SSL plug-ins}\label{sslModels}
\vskip 0.15in
\begin{center}
\begin{small}
\begin{minipage}{\columnwidth}
\begin{tabular*}{\textwidth}{@{\extracolsep{\fill}}lccccc@{\extracolsep{\fill}}}
\toprule%
{SSL plugin} & \multicolumn{2}{c}{Precision} & \multicolumn{2}{c}{Sensitivity} & \multicolumn{1}{c}{Trust} \\\cmidrule{2-3}\cmidrule{4-5}%
 & Pos. & Neg. & Pos. & Neg.  & \\
\midrule
MoCo (\textit{MIMIC-CXR}) & 0.995   & 0.896  & 0.884   & 0.995 & 0.909   \\
MoCo (\textit{ImageNet}) & 0.998   & 0.934  & 0.930   & 0.998 & 0.937   \\
SimCLR (\textit{ImageNet}) & \textbf{1.000}  &  \textbf{0.952} & \textbf{0.950}   & \textbf{1.000} & \textbf{0.954} \\
\bottomrule
\end{tabular*}
\end{minipage}
\end{small}
\end{center}
\vskip -0.1in
\end{table}

\section{Conclusion}
\label{conclusion}
In this work, we propose TRUDLMIA, a trustworthy and high-performing modular deep learning framework for medical image analysis, alleviating the prevalent issues of limited data, class imbalance, and low trustworthiness. The framework comprises large-scale supervised and self-supervised learning, as well as fine-tuning on downstream tasks in a supervised fashion. Through a highly successful assessment on the COVIDx dataset, TRUDLMIA framework proves its effectiveness for medical image analysis. \\ 

The proposed TRUDLMIA has shown great potential as a deep learning framework for medical image analysis due to its efficacy and simplicity. Models trained through the framework surpass traditional supervised models, including the state-of-the-art COVID-Net CXR-3. Our trustworthy model, COVID-CXR-SSL, will be made publicly available. We hope the TRUDLMIA can contribute to the ongoing fight against the pandemic and establish a viable path for future ones. \\

Our future work includes exploring the explainability of the models built using the framework, e.g., generating saliency maps with methods such as Grad-CAM \cite{gradcam}, in collaboration with radiologists. We are interested in investigating the use of the latest Vision Transformers \cite{vit} to replace the CNNs used in the TRUDLMIA framework. Furthermore, the use of Generative Adversarial Networks (GANs) \cite{ali2022combating} can be explored for augmenting dataset size, as well as its impact on model trust. 

\section{Potential Societal Impact}
Data collection for this study was conducted ethically from public and approved data. 
While promising, our work is by no means a production-ready solution. Thus, in all of our published files, a disclaimer is included, recommending prospective COVID-19 patients to seek help from professional medical practitioners.

\bibliographystyle{IEEEtran}  
\bibliography{references.bib}
\end{document}